\documentclass[11pt,preprint]{aastex}
\usepackage{graphicx,times,amsmath}
\newcommand{\ba}{\begin{eqnarray}}
\newcommand{\ea}{\end{eqnarray}}
\def\ra{\rightarrow}
\def\eps{\epsilon}
\def\vareps{\varepsilon}
\begin{document}

\title{Beta decay radiation signature from neutron-rich gamma-ray bursts?}

\author{Soebur Razzaque and Peter M\'esz\'aros}
\affil{Department of Astronomy \& Astrophysics, Department of Physics,
Pennsylvania State University, University Park, PA 16802}

\begin{abstract}
Core collapse of massive stars and binary neutron stars or black
hole-neutron star binary mergers are likely progenitors of long and
short duration gamma-ray bursts respectively. Neutronized material in
the former and neutron star material in the latter are ejected by the
central engine implying a neutron-rich jet outflow. A free neutron,
however, beta decays to a proton, an electron (beta) and an
anti-neutrino in about fifteen minutes in its rest frame. Sudden
creation of a relativistic electron is accompanied by radiation with
unique temporal and spectral signature. We calculate here this
radiation signature collectively emitted by all beta decay electrons
from neutron-rich outflow. Detection of this signature may thus
provide strong evidence for not only neutron but also for proton
content in the relativistic gamma-ray burst jets.
\end{abstract}


\section{Introduction}

The constituents of relativistic gamma-ray burst (GRB) jets are not
known completely as yet. The observed MeV energy $\gamma$-rays are
most likely radiated by energetic electrons which are shock
accelerated by internal collisions of successively ejected materials
by the GRB central engine. In this widely accepted {\em fireball shock
model}, protons along with electrons are needed to be present in the
GRB jet to explain the observed rapid variability of $\gamma$-rays and
the GRB afterglow. Free neutrons are speculated to be present in the
GRB jet \citep{dkk99a} which may be a common feature \citep{b03b}
following the ejection of neutronized core material in case of a long
burst which are most likely created by core collapse of massive stars,
a mechanism similar to that of the supernovae of type Ib/c or type
II. Recent observations of several short burst afterglows suggest a
common relativistic jet feature in both the long and short bursts. In
the latter case, neutron star material may also feed the jet with free
neutrons.

Observational signatures (both electromagnetic and neutrino) of a
neutron component in the GRB jets have been discussed by many authors,
in the early fireball evolution phase \citep{dkk99a, bm00, mr00, rm06}
and later in the internal shocks or afterglow phase \citep{dkk99b,
pd02, b03a, fw04, fzw05, da06}. However, most of these
predictions depend heavily on the jet model parameters and
interpretation of data may be difficult.

In this paper we discuss a generic signature of neutron-rich jets,
namely the radiation emitted by relativistic electrons created by
neutron beta decay: $n\ra p e^- {\bar \nu}_e$ in the GRB jets. This
radiation signature should be ubiquitiously present in the
electromagnetic signal detectable from all GRBs if the jet contains
free neutrons. We also find that the signal strength do not vary
rapidly for reasonable ratios of the neutrons to protons in the jet
and/or other jet parameters. Thus detection or non-detection of this
signature may serve as a powerful discriminator between GRB jets with
and without a neutron component.

We discuss the neutron-rich jet models of long and short bursts in
Section 2, beta decay electron energy and number distribution in
Section 3 and radiation signature in Section 4. We discuss our results
and detection prospects of beta decay radiation in Section 5.

\section{Neutron-rich jet models}

The typical parameters we use for long and short GRB jets are listed
in Table \ref{sample_parameters}. The baryon loading parameter (also
known as the dimensionless entropy) is defined as the total energy to
mass flow ratio in the GRB jet: $\eta=L/{\dot M}c^2$. The total mass
outflow rate in the jet, ignoring the electron mass ($m_e \ll m_p$)
and negligible thermal energy, is ${\dot M} \simeq 4\pi r^2 c
(1+\xi_o) n'_p m_p$. Here $\xi_o =n'_n/n'_p$ is the initial neutron to
proton number density ratio in the jet comoving frame. The outflow
starts at a radius $R_o$ which is a few times the Schwarzschild radius
$r_g = 2GM_{\rm bh}/c^2$ of a solar mass (few solar mass) black hole
for a short (long) GRB.

\begin{deluxetable}{lcc}
\tablewidth{0pt}
\tablecolumns{3}
\tablecaption{\label{sample_parameters}
Typical Gamma-Ray Burst Parameters}
\tablehead{
\colhead{Parameter} &
\colhead{Long GRB} &
\colhead{Short GRB} }
\startdata
Total isotropic-equivalent & $10^{52}L_{52}$ & $10^{50}L_{50}$ \\
energy outflow ($L$) ergs/s & & \\ Burst duration ($t_{90}$) s & 10 &
1 \\ Redshift ($z$) & 1 & 0.1 \\ Luminosity distance\tablenotemark{a}
($d_L$) cm & $10^{28}d_{L,28}$ & $10^{27}d_{L,27}$ \\ Baryon loading
parameter ($\eta$) & $316\eta_{2.5}$ & $316\eta_{2.5}$ \\ Initial
outflow radius ($R_o$) cm & $10^7R_{o,7}$ & $10^6R_{o,6}$
\enddata
\tablenotetext{a}{Using typical cosmological parameters}
\end{deluxetable}

Electrons, protons and neutrons are coupled to thermal radiation in
the expanding jet outflow as long as the Compton scattering time
scales $t'_{\rm Th} \simeq (n'_p \sigma_{\rm Th}c)^{-1}$ and elastic
$n$-$p$ scattering time scale $t'_{np} \simeq (n'_p
\sigma_{np}c)^{-1}$ are shorter than the plasma expansion time
$t'_{\rm exp} \simeq r/c\Gamma (r)$ in the jet comoving frame (see, e.g.,
\citet{rbr05} for a detailed description of the neutron-rich jet
dynamics). Here, $\Gamma(r)\propto r/R_o$ is the bulk Lorentz factor
of the expanding jet outflow and the elastic $n$-$p$ scattering
cross-section $\sigma_{np} \approx \sigma_{\rm Th}/20$, the Thomson
cross-section. The final value of $\Gamma$, after the expansion phase
is over, depends on the value of $\eta$. For sufficiently high value
of $\eta$, the neutrons may decouple from the outflow
\citep{dkk99a,bm00}. The corresponding critical $\eta$ value for
$n$-$p$ decoupling, from the condition $t'_{np} = t'_{\rm exp}$, is
\ba \eta_{np} \simeq \left[ \frac{L\sigma_{np}}{4\pi R_o m_p c^3
(1+\xi_o)} \right]^{1/4} \approx \begin{cases} 264
~(L_{52}/R_{o,7})^{1/4} (1+\xi_{o,1})^{-1/4} \cr 148
~(L_{50}/R_{o,6})^{1/4} (1+\xi_{o,1})^{-1/4}
\end{cases} \label{np-entropy} \ea
Here we have used $\xi_{o}= 10\xi_{o,1}$. The final bulk Lorentz
factor of the neutrons in the jet outflow is
\ba \Gamma_{n,f} = \begin{cases} R_{np}/R_o ~; & \eta >\eta_{np} \cr
\Gamma_{p,f} \approx \eta  ~; & \eta < \eta_{np} \end{cases} 
\label{gamma-final} \ea
where $R_{np}$ is the $n$-$p$ decoupling radius where the nuclear
scattering optical depth $\tau'_{np} \simeq n'_p \sigma_{np}
R_{np}/\Gamma(R_{np}) =1$ and is given by
\ba R_{np} &\simeq &
\eta_{np} \left( \frac{\eta_{np}}{\eta} \right)^{1/3}
\approx \begin{cases} 2.5\times 10^9 ~(L_{52} R_{o,7}^2/
\eta_{2.5})^{1/3} (1+\xi_{o,1})^{-1/3} ~{\rm cm} \cr 1.2\times 10^8
~(L_{50} R_{o,6}^2/ \eta_{2.5})^{1/3} (1+\xi_{o,1})^{-1/3} ~{\rm cm}
\end{cases} \label{np-radius} \ea
Note that for $\eta>\eta_{np}$ the final Lorentz factor for the
electron, proton and radiation outflow $\Gamma_{p,f} > \Gamma_{n,f}$
[see, e.g., \citet{rbr05}]. In any case $\Gamma_{n,f}\lesssim \eta$,
always.

The kinetic luminosity of the freely coasting neutron outflow in the
GRB jet is given by
\ba L_{n,\rm k} = L\frac{\Gamma_{n,f}}{\eta}
\frac{\xi_o}{1+\xi_o} \ea
Initially, at time $t\sim 0$, the neutron volume number density in the
outflow, in the jet comoving frame, is
\ba n'_{n, o} = \frac{L_{n,\rm k}}
{4\pi r^2\beta_n m_n c^3 \Gamma_{n,f}^2} \label{n-initial-dens} \ea
Here $\beta_n\sim 1$. However, this decreases exponentially with time
as neutrons decay ({\em beta decay}) which we discuss next.

\section{Beta decay}

A free neutron has a mean lifetime $\tau_{\beta} = 886.7$ s in its
rest frame. This corresponds to a mean decay radius $R_{\beta} =
c\tau_{\beta}\Gamma_{n,f}$ for a neutron at rest in the Lorentz
boosted outflow. The volume number density of beta decay electrons (or
protons or neutrinos) in the outflow evolves with time as $n'_e (t) =
n'_{n, o} (1-\exp[-r(t)/R_{\beta}])$ and the total number of beta
decay electrons at a time $t$ is $N'_e (t) = 4\pi r^2(t) n'_e(t)ct$
with no energy injection at later time $t\gtrsim 0$. We denote the
variables in the frame comoving with the neutron outflow with primes
from now on. The relationship between the observed time and neutron
outflow radius is $r(t)=2ct\Gamma_{n,f}^2/(1+z)$.

For our purpose we are interested in the total number of freshly
created electrons at a time $t$ which we define approximately, using
Equation~(\ref{n-initial-dens}), as
\ba {\cal N}'_e (t) \simeq \frac{{\rm min}[t,t_{90}] L_{n,\rm k}}
{m_n c^2 \Gamma_{n,f}^2} ~\times \begin{cases} 1-e^{-r(t)/R_{\beta}}
~; & r(t) \le R_{\beta} \cr e^{-r(t)/R_{\beta}} ~; & r(t) > R_{\beta}
\end{cases} \label{bdec-tot-e} \ea
The validity of this approximation depends on the fast rise and
exponential decay of the function with time. Here $t_{90}L_{n,\rm k}$
is the total energy carried by the neutron outflow in case of a GRB
central engine active for a time $t_{90}$. Note that in case of
complete decoupling of the neutron component from the rest of the
outflow for $\Gamma_{n,f} \ll \Gamma_{p,f}$, the observed GRB may
result from the leading proton outflow with $t_{90}$ duration and the
lagging neutron component may have a different time scale. However, we
ignore this possibility for the sake of simplicity.

The total energy released in a beta decay is the difference between
the rest mass energies of the initial and final particles:
$Q=m_nc^2-(m_p+m_e+m_{\nu})c^2 = 0.782 - m_{\nu}c^2 $ MeV [see, e.g.,
\citet{k87}]. This energy is shared by the kinetic energies of the
electron ($T_e$), neutrino ($T_{\nu}$) and proton ($T_p$) in the
decaying neutron's rest frame. With a negligible neutrino mass and
tiny proton recoil energy: $T_p \sim 0.3$ keV (because of its large
mass), the maximum electron's kinetic energy is $T_{e,\rm max} \simeq
Q$ for $T_{\nu} \ra 0$. The shape of the kinetic energy spectrum of
beta decay electrons is given by
\ba N_e(T_e) = \sqrt{T_e^2 + 2 T_e m_ec^2} (Q-T_e)^2 (T_e +m_ec^2)
\label{bdec-energy-dist} \label{bdec-spectrum} \ea
which vanishes for $T_e=0$ and $Q$.

\begin{figure}
\epsscale{1}
\plotone{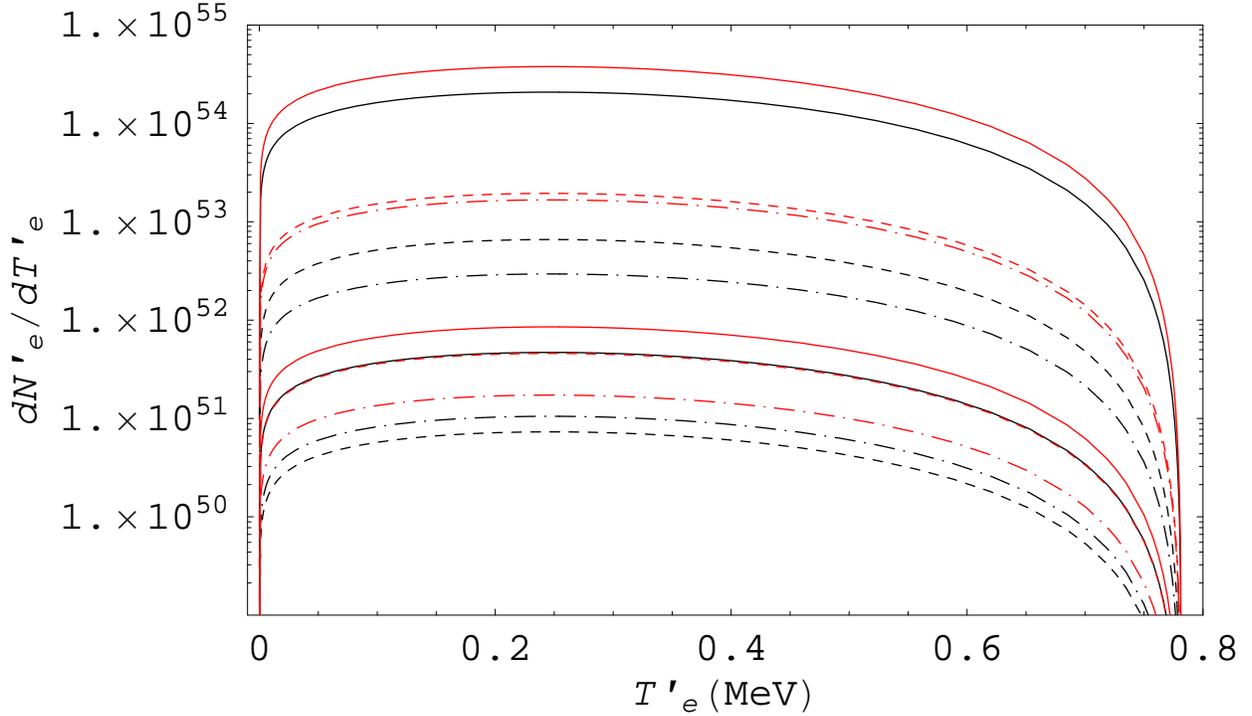}
\caption{\label{fig:ke-distribution}
Kinetic energy distribution [Equation (\ref{normalized-spectrum})] of
beta decay electrons normalized to the total number of electrons
[Equation (\ref{bdec-tot-e})] in the GRB jet comoving frame at an
observed time $t_{\beta}=\tau_{\beta}/2\Gamma_{n,f}$ for which
$r(t)=R_{\beta}$ or when the initial number of neutrons have decayed
by a factor $1/e$. The upper (lower) set of six curves correspond to
long (short) GRBs. The solid, dashed and dotted-dash pair of curves
correspond to $\eta = 100$, $316$ and $500$ with $\xi_o=10$
($\xi_o=1$) for the upper (lower) curve in each pair. The other
parameters for both the long and short GRBs are listed in Table
\ref{sample_parameters}. The calculated values of $\Gamma_{n,f}$ and 
$t_{\beta}$ for all curves are listed in Table
\ref{calculated_parameters}. }
\end{figure}

The kinetic energy of the electrons in Equation~(\ref{bdec-tot-e}) is
then distributed according to Equation~(\ref{bdec-spectrum}) and we
write a normalized kinetic energy distribution of beta decay electrons
at an observed time $t$, in the jet outflow, as
\ba \frac{d{\cal N}'_e}{dT'_e} (t) = \frac{{\cal N}'_e (t) N'_e(T'_e)}
{\int_{0}^{Q} N'_e(T'_e) dT'_e} \label{normalized-spectrum} \ea
We verify this expression by numerically integrating the right hand
side over $T'_e$ in the range $0$-$Q$ and comparing with Equation
(\ref{bdec-tot-e}) for different $t$. Figure \ref{fig:ke-distribution}
shows the distribution in Equation (\ref{normalized-spectrum}) plotted
for different parameters for long and short GRB jets at a time
$t_{\beta}=\tau_{\beta}/2\Gamma_{n,f}$ (assuming $z\sim 0$) when the
total number of freshly created beta decay electrons in Equation
(\ref{bdec-tot-e}) is the maximum. Note that ${\cal N}'_e$ increases
with $\xi_o$, however, it is not directly proportional since
$\Gamma_{n,f}$ is not directly proportional to $\xi_o$ [see
Equation~(\ref{gamma-final})]. We have listed different values of
$t_{\beta}$ and $\Gamma_{n,f}$ in Table
\ref{calculated_parameters} for different choice of $\eta$ and $\xi_o$ 
in case of long and short burst curves plotted in Figure
\ref{fig:ke-distribution}.

\section{Radiation spectrum and lightcurve}

Beta decay electrons are created suddenly with a peak kinetic energy
of $0.25$ MeV in the neutron's rest frame. These electrons may be
thought to be created initially at rest and then accelerated to a
constant high speed $v=c\beta_e$ within a short time interval $\tau
\sim \hbar c/E_e$, following the uncertainty principle. Here $E_e =
T_e +m_ec^2$ is the total energy of the electron. Rapidly accelerating
electrons then emit radiation (often called {\em inner bremsstrahlung}
in the literature; see \citet{pr72} for an early astrophysical
application) similar to the bremsstrahlung. The radiation spectrum
(total energy radiated per unit frequency interval) is also flat as
bremsstrahlung and is given by \citep{j99}
\ba \frac{dI}{d\omega} = \frac{q^2}{\pi c}
\left[ \frac{1}{\beta_e} {\rm ln}\left( \frac{1+\beta_e}{1-\beta_e}
\right) -2 \right] \label{radiation-spectrum} \ea
where the electron's speed (with momentum $p_e$) is
\ba \beta_e = \frac{p_ec}{E_e} = \frac{\sqrt{T_e^2 + 2 T_e m_ec^2}}
{T_e +m_ec^2} \approx 0.74 \label{electron-speed} \ea
The last number is for electrons of peak kinetic energy. Note that the
spectrum defined in Equation (\ref{radiation-spectrum}) is
proportional to the more familiar notation $F_{\nu}$.

The maximum energy of the emitted photons from beta decay electrons is
$\hbar\omega_{\rm max} \lesssim E_e$ following the condition
$\omega\tau\lesssim 1$, although a full quantum mechanical treatment
is necessary to evaluate the precise value. Semi-classically an
electron loses a small fraction of its energy given by integrating
Equation (\ref{radiation-spectrum}) over $\omega$ as
\ba \frac{E_{\gamma}}{E_e} = \frac{q^2}{\pi \hbar c}
\left[ \frac{1}{\beta_e} {\rm ln}\left( \frac{1+\beta_e}{1-\beta_e}
\right) -2 \right] \approx 0.001 \label{radiation-loss} \ea
Again, the last number is for an electron of peak kinetic energy 0.25
MeV. However, the collective radiation energy from all beta decay
electrons in a GRB jet is large. For an observed GRB, the beta decay
radiation is beamed along the line of sight of an observer because of
the relativistic bulk motion of the jet even though beta decay
electrons do not have any particular angular orientation and the
radiation is emitted along the electron's directon in the jet comoving
frame.

We write the total energy radiated per unit frequency by all beta
decay electrons at a time $t$ in Equation (\ref{normalized-spectrum})
as would be measured by an observer as
\ba \frac{d{\cal I}}{d\omega} (t)  &=&
\frac{\Gamma_{n,f}}{(1+z)}
\int_{0}^{Q} \frac{d{\cal N}'_e}{dT'_e}(t) \frac{dI'}{d\omega'} dT'_e
\nonumber \\ &=&
\frac{q^2\Gamma_{n,f}} {\pi c (1+z)}
\int_{0}^{Q} \frac{d{\cal N}'_e}{dT'_e}(t)
\left[ \frac{1}{\beta'_e} {\rm ln}\left(
\frac{1+\beta'_e}{1-\beta'_e} \right) -2 \right] dT'_e
\label{bdec-radiation-spectrum} \ea
which is flat in frequency. The corresponding total radiation energy
emitted over all frequencies at a time $t$, by integrating Equation
(\ref{bdec-radiation-spectrum}) over $0 \lesssim \hbar\omega \lesssim
(T'_e +m_ec^2) \Gamma_{n,f}/(1+z)$, is
\ba {\cal E}_{\gamma,\rm tot} (t) &\simeq &
\frac{q^2\Gamma_{n,f}^2}{\pi\hbar c
(1+z)^2} \int_{0}^{Q} (T'_e +m_ec^2) \frac{d{\cal N}'_e}{dT'_e} (t)
\left[ \frac{1}{\beta'_e} {\rm ln}\left(
\frac{1+\beta'_e}{1-\beta'_e} \right) -2 \right] dT'_e
\label{total-energy} \ea

\begin{figure}
\epsscale{1}
\plotone{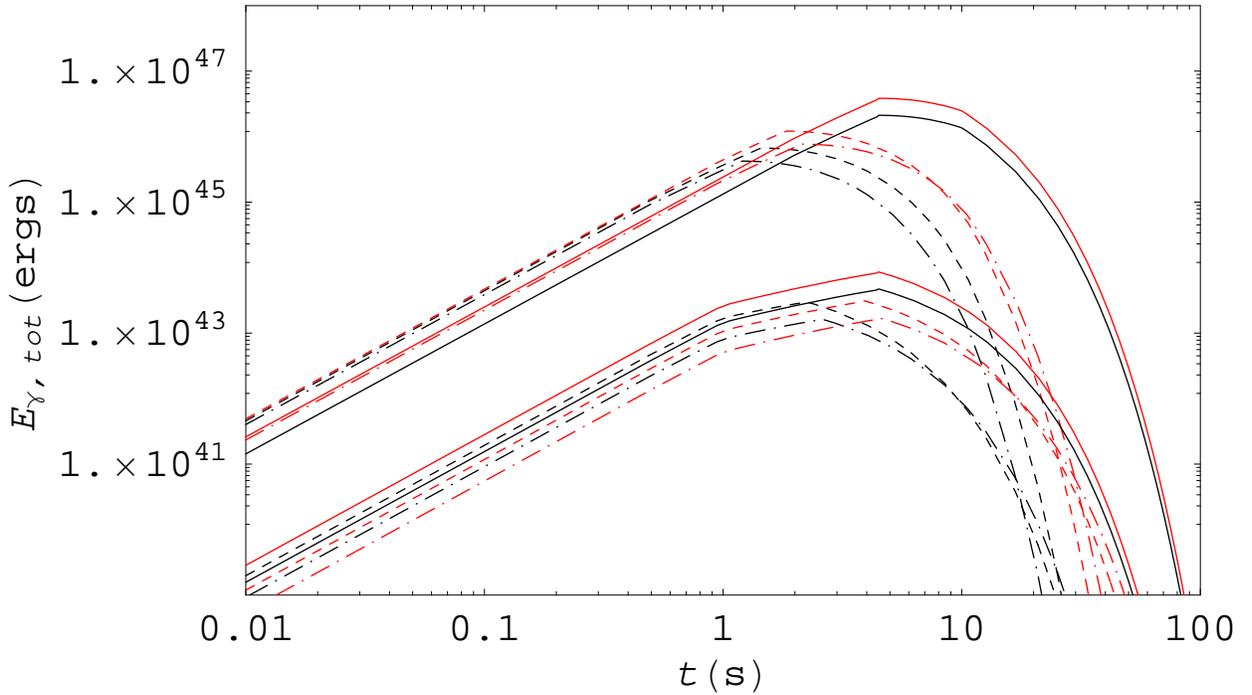}
\caption{\label{fig:total-energy}
Total radiation energy emitted by beta decay electrons over all
frequencies as function of time [Equation (\ref{total-energy})]. The
model parameters for the GRB jets for different curves are the same as
in Figure \ref{fig:ke-distribution} and listed in Table
\ref{calculated_parameters}. The upper (lower) set of curves
correspond to the long (short) bursts. Note that the signal is
stronger and peaks later for $\Gamma_{n,f} = \eta$ cases as evidenced
by the solid pair of curves in case of both long and short bursts. The
dependence on $\xi_o$ is rather weak.}
\end{figure}

We have plotted the total energy radiated by beta decay electrons in
Figure \ref{fig:total-energy} as function of observed time by
numerically evaluating the integral in Equation (\ref{total-energy})
for different GRB model parameters used in Figure
\ref{fig:ke-distribution} and listed in Table
\ref{calculated_parameters}. The total number of electrons at the peak
kinetic energy plotted in Figure \ref{fig:ke-distribution} times the
total energy radiated by each of these electrons using Equation
(\ref{radiation-loss}) roughly corresponds to the peak total energy
plotted in Figure \ref{fig:total-energy} after multiplying with the
corresponding $\Gamma_{n,f}^2$ factor. The maximum observed photon
energy in all cases is $\eps_{\gamma,\rm max} \approx
1.3\Gamma_{n,f}/(1+z)$ MeV.

\section{Detection prospects and Discussion}

The highest energy photons carry most of the radiation energy plotted
in Figure \ref{fig:total-energy}, since the beta decay radiation
spectrum from Equation (\ref{bdec-radiation-spectrum}) is flat.  The
corresponding photon flux (energy received per unit area per unit
time) on Earth, in a given energy band
$\eps_{\gamma,l}$-$\eps_{\gamma,u}$; is
\ba {\cal F} ({\eps_{\gamma,l}} {\rm -} \eps_{\gamma,u})  
\simeq \frac{1}{4\pi d_L^2} \int_{\eps_{\gamma,l}}^{\eps_{\gamma,u}} 
\frac{d{\cal I}}{d\omega} (t) \label{beta-decay-flux} \ea
Upcoming high energy $\gamma$-ray detectors such as the {\em Large
Area Telescope} (LAT) on board the {\em Gamma-ray Large Area
Telescope} (GLAST) are sensitive to photons in the energy range $20$
MeV - $300$ GeV. The threshold fluence for LAT is $\sim 4\times
10^{-8}$ ergs/cm$^2$ for a short integration time. This roughly
corresponds to a few photons with energy $\gtrsim 0.1$ GeV collected
within a few seconds in its average $5000$ cm$^2$ effective area over
this energy range. We have listed in Table \ref{calculated_parameters}
the expected beta decay radiation flux at photon energies
$\eps_{\gamma} \gtrsim 0.1$ GeV, for different values of the relevant
parameters. We used a luminosity distance $d_L = 10^{27}d_{L,27}$ cm
corresponding to a redshift $z\sim 0.1$ for both the long and short
bursts. This redshift is typical for short bursts, and is at the lower
end for long bursts. (Observations indicate an occurrence rate of
$\sim 0.25$ year$^{-1}$ for long bursts and $2$-$3$ year$^{-1}$ for
short bursts in this redshift range).

\begin{deluxetable}{llccc}
\tablewidth{0pt}
\tablecolumns{5}
\tablecaption{\label{calculated_parameters}
Beta Decay Radiation Parameters}
\tablehead{
\colhead{Parameter} &
\colhead{} &
\colhead{$\Gamma_{n,f}$} &
\colhead{$t_{\beta}$ (s)} &
\colhead{${\cal F}$\tablenotemark{a}} }
\startdata
\cutinhead{long GRB}
$\eta = 500$ & $\xi_0=1$ & 376 & 1.2 & 0.33 \\
             & $\xi_o=10$ & 213 & 2.1 & 0.61 \\
$\eta = 316$ & $\xi_0=1$ & 316 & 1.4 & 0.53 \\
             & $\xi_o=10$ & 248 & 1.8 & 0.97 \\
$\eta = 100$ & $\xi_0=1$ & 100 & 4.4 & 1.7 \\
             & $\xi_o=10$ & 100 & 4.4 & 3.1 \\
\cutinhead{short GRB}
$\eta = 500$ & $\xi_0=1$ & 174 & 2.5 & 2.6 \\
             & $\xi_o=10$ & 99 & 4.5 & 1.4 \\
$\eta = 316$ & $\xi_0=1$ & 203 & 2.2 & 2.4 \\
             & $\xi_o=10$ & 115 & 3.8 & 4.9 \\
$\eta = 100$ & $\xi_0=1$ & 100 & 4.4 & 3.8 \\
             & $\xi_o=10$ & 100 & 4.4 & 6.9 \\
\enddata
\tablenotetext{a}{in units of $(\times 10^{-9})$ 
ergs cm$^{-2}$ s$^{-1}$ for long GRBs 
and in units of $(\times 10^{-12})$ 
ergs cm$^{-2}$ s$^{-1}$ for short GRBs}
\end{deluxetable}

However, the beta decay signal can be diluted by other types of
emission in the same GRB, since the signal can coincide temporally
with the usual prompt $\gamma$-ray emission phase. This is somewhat
mitigated by the fact that the two signals have different spectra, and
the beta decay signal has a smoother rise and quick decay, compared to
the sometimes erratic and longer signal of the long burst prompt
emission.  The GRB afterglow emission occurs on an even longer time
scale than the beta decay time scale $t_{\beta}$, and should not
interfere with the latter.  Hence we discuss the GRB prompt
$\gamma$-ray emission as a source of background radiation below.

The observed isotropic-equivalent bolometric $\gamma$-ray luminsosity,
$L_{\gamma}$, of a long (short) GRB is $\sim 10^{51}L_{\gamma,51}$
ergs/s ($\sim 10^{49}L_{\gamma,49}$ ergs/s) which is mostly
concentrated at the peak energy $\eps_{\gamma,\rm pk} =
100\eps_{\gamma,-4}$ keV. In the {\em fireball shock model}, this
energy corresponds to a fraction $\vareps_{e} \simeq L_{\gamma}/L \sim
0.1\vareps_{e,-1}$ of the total energy converted by shock accelerated
electrons. The luminosity at higher $\eps_{\gamma}$ decreases by a
power-law, with index $2\lesssim \alpha \lesssim 3$, following the
synchrotron and/or inverse Compton (IC) radiation spectrum by a
power-law distributed electrons. For $\alpha>2$, most of the energy in
a given band $\eps_{\gamma,l}$-$\eps_{\gamma,u}$ is carried by the low
energy photons (unlike in the beta decay signal). The corresponding
prompt (background) synchrotron/IC photon flux at Earth is
approximately
\ba F_{\rm prompt} (\eps_{\gamma,l}-\eps_{\gamma,u}) 
&\simeq & \frac{L_{\gamma}} {4\pi d_L^2} \left( \frac{\eps_{\gamma,l}}
{\eps_{\gamma,\rm pk}}
\right)^{2-\alpha} ~;~ \eps_{\gamma,l} \gtrsim \eps_{\gamma,\rm pk}
\nonumber \\ &\approx & 8\times 10^{-8}
\left( \frac{\eps_{\gamma,l}} {0.1~{\rm GeV}} \right)^{-1} 
\frac{L_{\gamma, 51}} {d_{27}^{2} \eps_{\gamma,-4}}
 ~{\rm ergs}~{\rm cm}^{-2} ~{\rm s}^{-1} ~;~ \alpha=3
\label{synchroton-flux} \ea
for a long GRB at $z\sim 0.1$. Thus, the detection of the beta decay
radiation signature would be background limited in the case of long
bursts with $t_{90} \gtrsim 2$ s, if the prompt spectrum extends to
$\sim 0.1$ GeV with the above nominal spectrum. However, one can
expect a sizable fraction of long GRBs where the prompt spectrum does
not contain significant energy at 0.1 GeV, e.g. due to a spectral
index which is steeper than the nominal value $\alpha=3$, or due to a
fall-off in the spectrum [e.g., \citet{pretal00}; also, as shown by
\citet{r04} that a fraction of long GRBs are dominated by a thermal-like
component, with little or no emission above a few MeV].  While data in
the $\sim 0.1$ GeV range is currently still limited, one may
conservatively estimate at $30\%$ level the fraction of all long
GRBs which do not have substantial emission at 0.1 GeV, and hence
which contribute no background for the beta decay signal.

On the other hand, for short bursts, the beta decay signature peaks on
timescales longer (see Figure \ref{fig:total-energy}) than the prompt
emission ($t_{90}\lesssim 2$ s), and the beta decay signal is expected
to be essentially free from a prompt emission background at 0.1 GeV.

One other source of background to consider is the diffuse
extragalactic $\gamma$-ray background measured by EGRET
\citep{egret97}. The corresponding photon number flux may be fitted in
the $30$ MeV-$120$ GeV energy range as $dN/d\eps_{\gamma} \approx
7.32\times 10^{-9} (\eps_{\gamma}/451~{\rm MeV})^{-2.1}$ cm$^{-2}$
s$^{-1}$ sr$^{-1}$ MeV$^{-1}$. This background energy flux within
$\sim 1^{\circ}$ angular resolution of LAT is $F_{\rm bkg} \approx
3\times 10^{-14}$ ergs cm$^{-2}$ s$^{-1}$ for $\eps_{\gamma}\gtrsim
0.1$ GeV which is negligible compared to the beta decay flux listed in
Table \ref{calculated_parameters} or GRB prompt flux in Equation
(\ref{synchroton-flux}).

The predicted fluxes ${\cal F} (\gtrsim 0.1 ~\rm GeV)$ in Table
\ref{calculated_parameters} were calculated for bursts  of nominal (MeV)
luminosities $L_\gamma=10^{51}$ erg/s (long bursts) and
$L_\gamma=10^{49}$ erg/s (short bursts), at a redshift
$z=0.1$. Multiplying these nominal fluxes by a typical average beta
decay peak duration $t_\beta\sim 5$ s, and comparing to the LAT
threshold sensitivity fluence $4\times 10^{-8}$ ergs/cm$^{2}$ at
$\eps_{\gamma}\sim 0.1$ GeV, one sees that for short bursts, even
though it is background free, the signal is undetectable by the
LAT. For long bursts, however, it may be possible to detect the beta
decay radiation signature from nearby long bursts whose luminosity is
ten times the average value.  Bursts satisfying these criteria may
represent $\sim 0.1$-$1\%$ of a nominal total of 100 bursts per year
detected by GLAST. This suggests that GLAST may be marginally able to
detect such signals, and future generations of very energetic
space-based $\gamma$-ray large area telescopes may be able to
quantitatively explore this problem.

Non-thermal multi GeV energy neutrino and $\gamma$-ray emission from
the $n$-$p$ decoupling phase can indicate the presence of a neutron
component in the GRB jets, for values of the dimensionless entropy
$\eta \gtrsim \eta_{np}$, which is typically an unknown parameter.  On
the other hand, the electromagnetic radiation signature discussed here
is valid for all values of $\eta$. The evidence for a neutron
component from the beta decay radiation signature would definitely
imply the presence of protons in the GRB jet, since neutrons are
coupled to them at least initially. Even though the converse is not
true, the beta decay radiation signature may be an alternate way to
explore the baryon loading in the GRB jets in the cases when high
energy neutrinos from the internal or external shocks are absent
(e.g., if the protons are not co-accelerated with electrons), or if
too few are detected. The beta decay photon signature of neutron decay
may thus be a valuable tool for investigating the particle
acceleration process and possibly constrain the progenitors of
Gamma-Ray Burst sources.

\acknowledgements{Research supported in part through NSF AST 0307376
and NASA NAG5-13286.}

\end{document}